# Observation of magnetoelectric behavior at room temperature in Pb(Fe$_{1-x}$Ti$_x$)O$_3$

V. R. Palkar[*] and S. K. Malik

Tata Institute of Fundamental Research, Mumbai 40005, India

The coexistence of ferroelectric and ferromagnetic properties at room temperature is very rarely observed. We have been successful in converting ferroelectric PbTiO$_3$ into a magnetoelectric material by partly substituting Fe at the Ti site. The Pb(Fe$_{1-x}$Ti$_x$)O$_3$ system exhibits ferroelectric and ferromagnetic ordering at *room temperature*. Even more remarkably, our results demonstrate a coupling between the two order parameters. Hence it could be a futuristic material to provide cost effective and simple path for designing novel electromagnetic devices.

[*]corresponding author e-mail address: palkar@tifr.res.in

Ferroelectric materials have wide applications such as in non-volatile memories, capacitors, transducers, actuators, etc. Likewise, magnetic materials are also used in data storage and a variety of other applications. There is great technological promise and fundamental interest if ferroelectricity and ferromagnetism coexist at *room temperature*. Further, if there is a coupling between the two order parameters, it should be possible to cause electric polarization by applying either an electric or a magnetic field. Thus writing of a data bit with an electric field and reading it with a magnetic field, and vice-versa, will be possible. This certainly offers additional degree of freedom in device designing. However, in reality, there is a scarcity of materials exhibiting magnetoelectric behavior at room temperature [1] possibly due to the fact that transition metal d electrons, which are essential for the presence of magnetic moment, reduce lattice distortion which is essential for ferroelectric behavior. Thus, additional structural or electronic driving force is required for ferroelectric and magnetic ordering to coexist [2]. Suchtelen *et al* [3] suggested the realization of composites of piezoelectric and magnetostrictive phases which could be electromagnetically coupled via stress mediation. Various groups have therefore, focused efforts on realization of bulk composites or heterostructural layers [4-7]. However, it is important to note that the structural non-compatibility and reactivity between two materials and also with the substrate, generates immense difficulties in growing heterostructures and achieving coupling between the two order parameters. Hence search for new materials exhibiting both these phenomena at room temperature continues.

Recently, we have shown that Tb doping at the Bi site induces ferrimagnetism in antiferromagnetic $BiFeO_3$ without disturbing ferroelectric ordering [8-9]. Here we report the conversion of ferroelectric PbTiO$_3$ to a compound with even superior magnetoelectric properties by partially substituting Ti with Fe. Remarkably, both magnetic ($T_M$) and ferroelectric transitions ($T_C$) occur above room temperature in this newly synthesized Pb(Fe$_x$Ti$_{1-x}$)O$_3$ system. The observed coupling between the two order parameters is an added advantage. The new material carries more importance since the basic compound, (PbTiO$_3$), has been thoroughly studied for non-volatile memories and other applications. So the related knowledge available in the literature would act as a guide while designing novel devices from this new compound.

Lead Titanate (PbTiO$_3$) is a well-studied displacive type of ferroelectric material belonging to the ABO$_3$-type perovskite group of compounds. It is well understood that Ti(3d)-O(2p) hybridization helps in stabilizing the ferroelctric distortion in PbTiO$_3$ [10]. Ferroelectric polarization mainly results from displacement of Ti$^{4+}$ ions with respect to the oxygen cage. Hybridization of the lowest energy level state with O(2p) states seems to be a requirement for stabilization of the ferroelectric state. The reduction in lattice distortion, after doping with transition metal element at Ti site, is expected as described by Hill [2]. However, PbTiO$_3$ is known for its very large lattice distortion (*c/a* ~1.064). It was, therefore, thought that the lattice could still sustain its tetragonal structure to some extent even after substituting transition metal elements like Fe at Ti site thus making it possible to induce magnetism in PbTiO$_3$ without disturbing its ferroelectric behavior. With this in mind, Pb(Fe$_x$Ti$_{1-x}$)O$_3$ samples, with varying x, were synthesized and their properties were investigated. The goal was to achieve a magnetoelectric material with reasonably good properties.

Co-precipitation method was developed to synthesize Pb(Fe$_x$Ti$_{1-x}$)O$_3$ samples so as to gain better control on chemical homogeneity and stoichiometry. Stoichiometric solution of lead (Pb$^{2+}$), titanium (Ti$^{4+}$) and ferric (Fe$^{3+}$) nitrates was initially prepared All three cations were then co-precipitated using NH$_4$OH as a precipitating agent. The hydroxy complex formed during precipitation was calcined at 700°C to decompose into well crystalline reacted phase. The reacted material was then characterized and studied for different properties. Powder X-ray diffraction patterns were obtained (Jeol X-ray diffractometer) using Cu K$_\alpha$ radiation. AC magnetization was measured using a home-built set up while dc magnetization was measured using a SQUID magnetometer (MPMS, Quantum Design). Dielectric response as a function of temperature was obtained from capacitance measured using HP LCZ meter at frequency of 1MHz. Ferroeletric hysteresis loop measurements were carried out on pellet sample sintered at 1000 °C/ 2 hours with silver as top and bottom electrode using Ferroelectric Loop Tracer (RT66).

From XRD pattern (Fig. 1), it is clear that even after substituting Ti with 50 mol % of Fe, the crystal structure remains tetragonal like that of the parent compound (PbTiO$_3$). XRD indicated the presence of a very small amount of pyrochlore (Pb$_2$Ti$_2$O$_6$) phase as an impurity in the sample. The observed change in lattice parameters is an indication of the formation of solid solution. As the amount of Fe increases, the tetragonal distortion (*c/a*) decreases as per expectation[2]. For Pb(Fe$_{0.5}$Ti$_{0.5}$)O$_3$ sample, lattice parameter *a* increases from 3.899Å to 3.928 Å while lattice parameter *c* decreases from 4.15 Å to 4.036 Å. As a result, tetragonal ratio (*c/a*) reduces from 1.064 to ~1.028. It was obvious that additional increase in Fe content would lead to further suppression in lattice distortion and hence the ferroelectricity. Therefore, further studies were concentrated on Pb(Fe$_{0.5}$Ti$_{0.5}$)O$_3$ composition.

High temperature ac-scusciptibility ($\chi_{ac}$) measurement on Pb(Fe$_{0.5}$Ti$_{0.5}$)O$_3$ sample, using a field of 1.1 Oe at 245 Hz, showed a peak at ~270°C (Fig. 2). Dielectric constant and loss factor, tan δ, as a function of frequency are shown in figure 3. Both show decrease with increase in frequency at initial stage. However, at higher frequencies dielectric constant gets more or less saturated. Hence dielectric measurements as a function of temperature were carried out at 1MHz (Fig. 4) which revealed the presence of a broad peak around 260°C. In magnetoelectrically ordered systems, dielectric anomaly near magnetic transition is predicted by Landau-Devonshire theory of phase transition [11]. Such an anomaly arises due to the influence of vanishing magnetic order on the electric order. Therefore, the peak obtained in $\chi_{ac}$ at ~270°C (Fig. 2) could be attributed to magnetic transition. In addition, dielectric response measurements exhibited a second peak around 420°C in Pb(Fe$_{0.5}$Ti$_{0.5}$)O$_3$ sample (Fig. 4). This peak could be assigned to the ferroelectric transition. Pure PbTiO$_3$ is known to have a ferroelectric transition (T$_C$) at ~490°C. With Fe substitution, the transition temperature is expected to decrease as a consequence of reduction in lattice distortion from 1.064 to 1.028. Though a small hump in loss factor (tan δ) has been observed near 200°C, it shoots up sharply at higher temperatures and goes out of measurement range (Fig. 4).

Magnetization–Field (M-H) isotherm for Pb(Fe$_{0.5}$Ti$_{0.5}$)O$_3$, at 300K (Fig. 5) shows saturation in M consistent with the ferromagnetic nature of Pb(Fe$_{0.5}$Ti$_{0.5}$)O$_3$ sample at room temperature. Saturated ferroeletric hysteresis loop (Fig. 6), obtained at room temperature proves the presence of ferroelectricity in the same sample.

In order to look for the coupling between the two order parameters in Pb(Fe$_{0.5}$Ti$_{0.5}$)O$_3$, ferroelectric loop measurements were performed initially on as-sintered pellet and then after poling the same pellet by applying a DC magnetic field of 2 Tesla. The coupling is evident from the effect seen on the saturation polarization (P$_s$) in magnetically poled sample (Fig. 5). As sintered sample does not give clear saturation whereas, magnetically poled sample shows saturation as well as an increase in polarization value. This implies the alignment of ferroelectric domains by means of an applied magnetic field. Further, the change in net magnetization was determined after poling the as-sintered sample by means of DC electric field (10kV/cm). Effectively, 10-12% enhancement in magnetization was detected. These results clearly demonstrate that a coupling between the two order parameters exists in this system.

In summary, we are successful in synthesizing a new compound with magnetoelectric properties at *room temperature*. It is significant that the co-existence occurs in a stable perovskite structure which will allow a clear search path for more new materials among this family of compounds. For example changing the doping element is a likely route in this direction. Moreover, the results point to a simple materials approach to novel device principles. Besides potential applications, detailed theoretical and experimental studies of this class of systems will hopefully lead to a rich and fascinating fundamental physics.

**Acknowledgement**

We appreciate guidance and encouragement received from Prof. S. Bhattacharya. We thank Prof. P. L. Paulose for his valuable co-operation. We are also thankful to B.A. Chalke and D. Buddhikot for experimental help.

**Figure Captions:**

**Figure 1** X-ray diffraction pattern of $Pb(Fe_{0.5}Ti_{0.5})O_3$ sample. Lines marked with '*' are from $Pb_2Ti_2O_6$ impurity phase.

**Figure 2** AC Magnetization vs. Temperature ($M_{ac}$-T) for $Pb(Fe_{0.5}Ti_{0.5})O_3$ (magnetic field of 1.1 Oe at 245 Hz).

**Figure 3.** Dielectric Constant and tan δ vs. Frequency at room temperature for $Pb(Fe_{0.5}Ti_{0.5})O_3$ pellet sintered at 1000ºC/2 hour.

**Figure 4** Dielectric Constant and tan δ vs. Temperature for $Pb(Fe_{0.5}Ti_{0.5})O_3$ pellet sintered at 1000ºC/2 hour. Measurements were done at 1MHz.

**Figure 5.** Magnetization vs. Field (M-H) isotherm for $Pb(Fe_{0.5}Ti_{0.5})O_3$ sample at 300 K

**Figure 6.** Ferroelectric Hysteresis Loop of $Pb(Fe_{0.5}Ti_{0.5})O_3$ pellet sample, a) as sintered, b) poled at 2 Tesla of DC Magnetic Field.

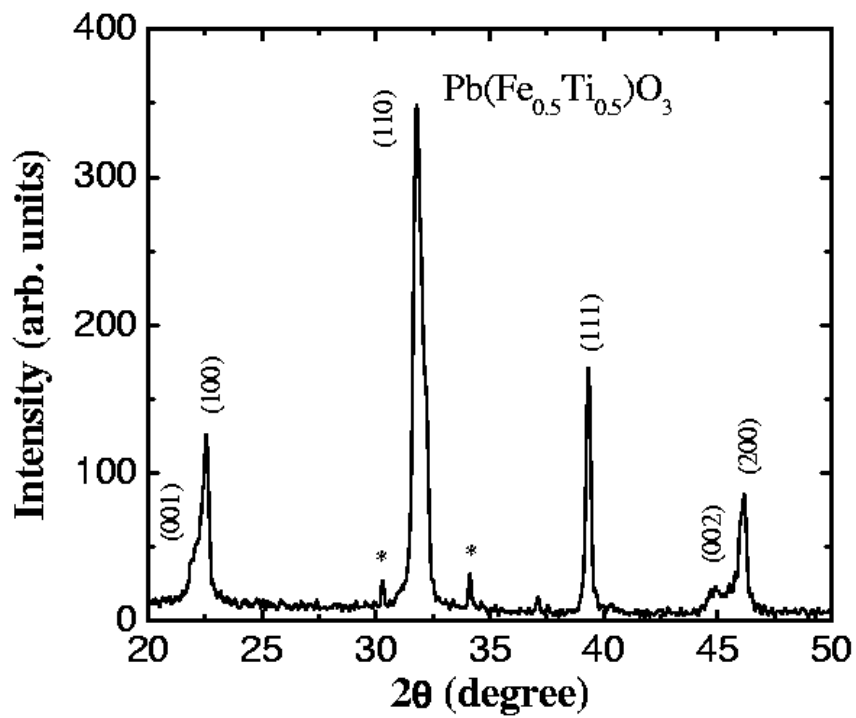

Figure 1. Palkar *et al.*

**Figure 1** X-ray diffraction pattern of Pb(Fe$_{0.5}$Ti$_{0.5}$)O$_3$ sample. Lines marked with '*' are from Pb$_2$Ti$_2$O$_6$ impurity phase.

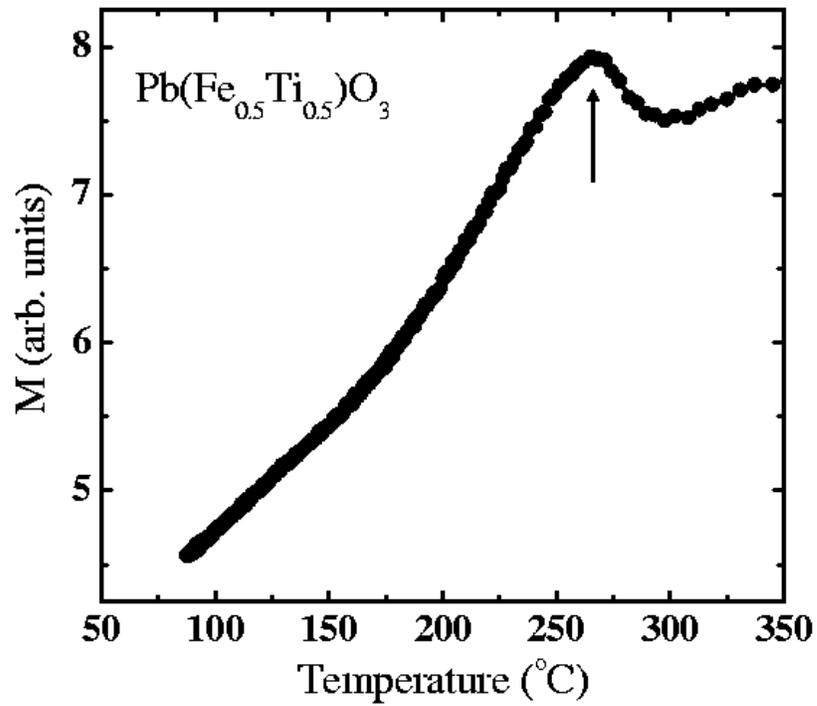

Figure 2. Palkar *et al.*

**Figure 2** AC Magnetization vs. Temperature ($M_{ac}$-T) for $Pb(Fe_{0.5}Ti_{0.5})O_3$ (magnetic field of 1.1 Oe at 245 Hz).

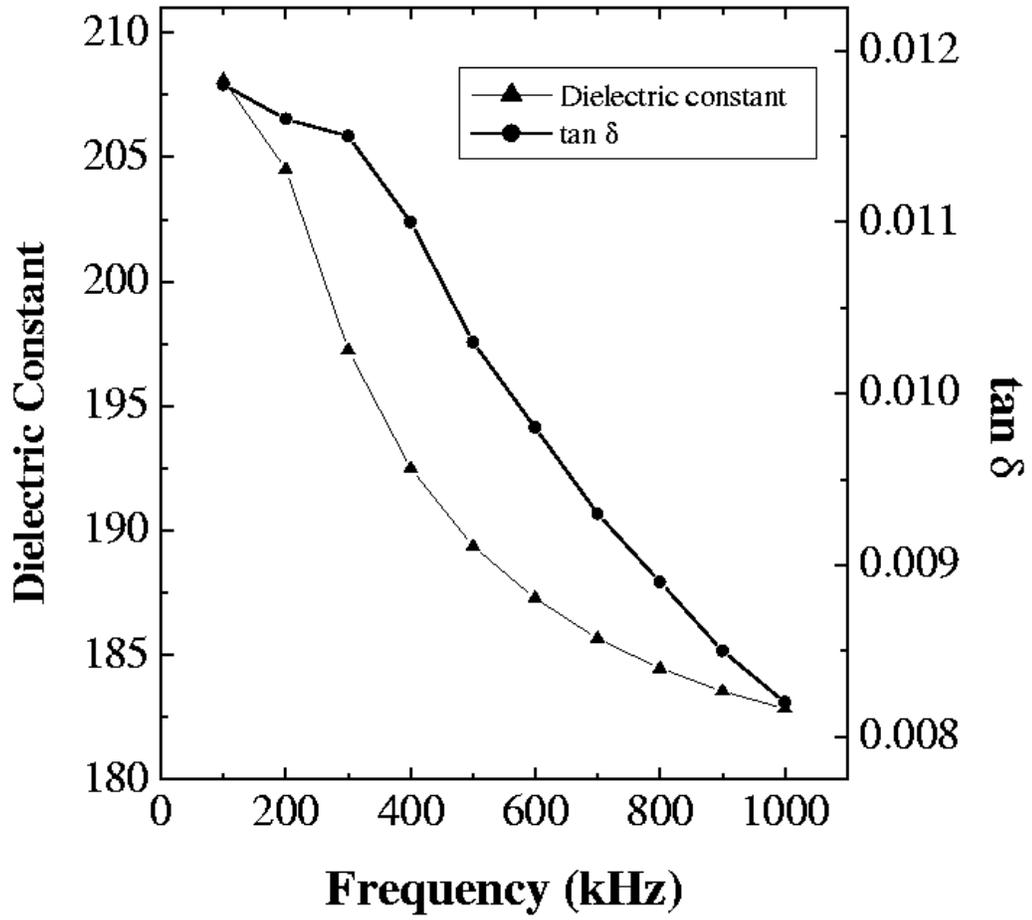

Figure 3. Palkar *et.al*

**Figure 3.** Dielectric Constant and tan δ vs. Frequency at room temperature for Pb(Fe$_{0.5}$Ti$_{0.5}$)O$_3$ pellet sintered at 1000°C/2 hour.

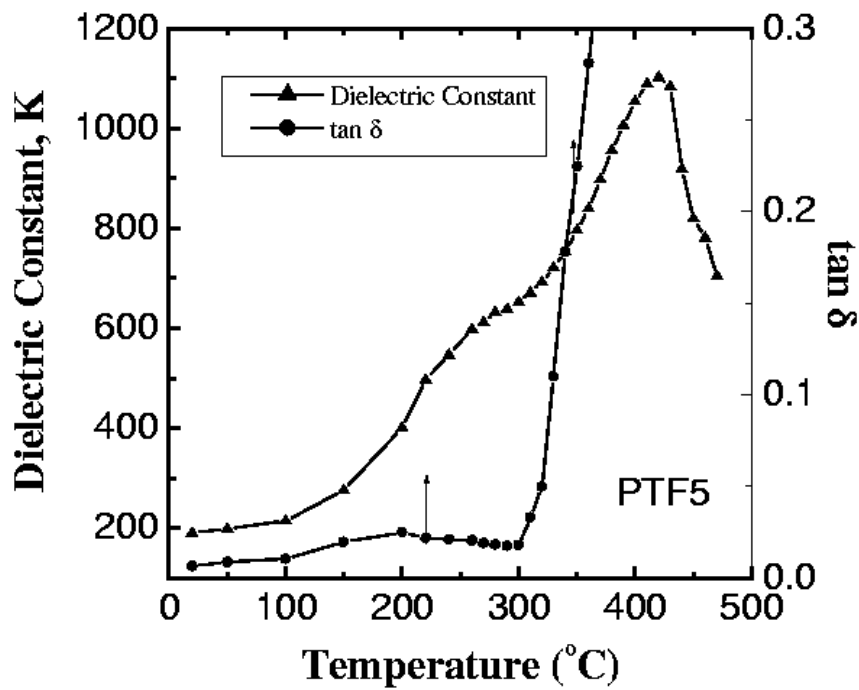

Figure 4. Palkar *et.al*

**Figure 4** Dielectric Constant and tan δ vs. Temperature for Pb(Fe$_{0.5}$Ti$_{0.5}$)O$_3$ pellet sintered at 1000ºC/2 hour. Measurements were done at 1MHz.

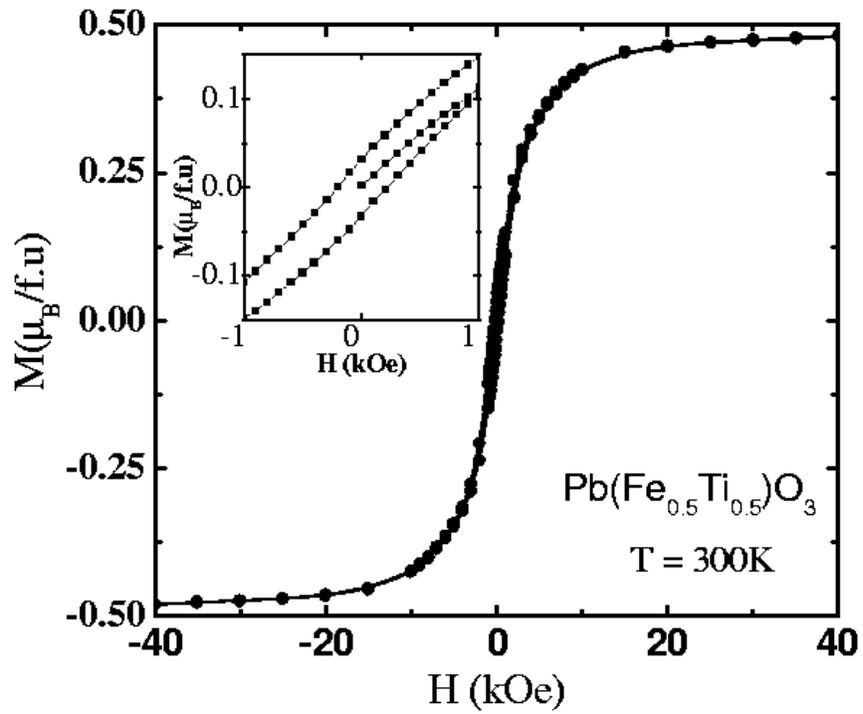

Figure 5. Palkar *et al.*

**Figure 5.** Magnetization vs. Field (M-H) isotherm for Pb(Fe$_{0.5}$Ti$_{0.5}$)O$_3$ sample at 300 K

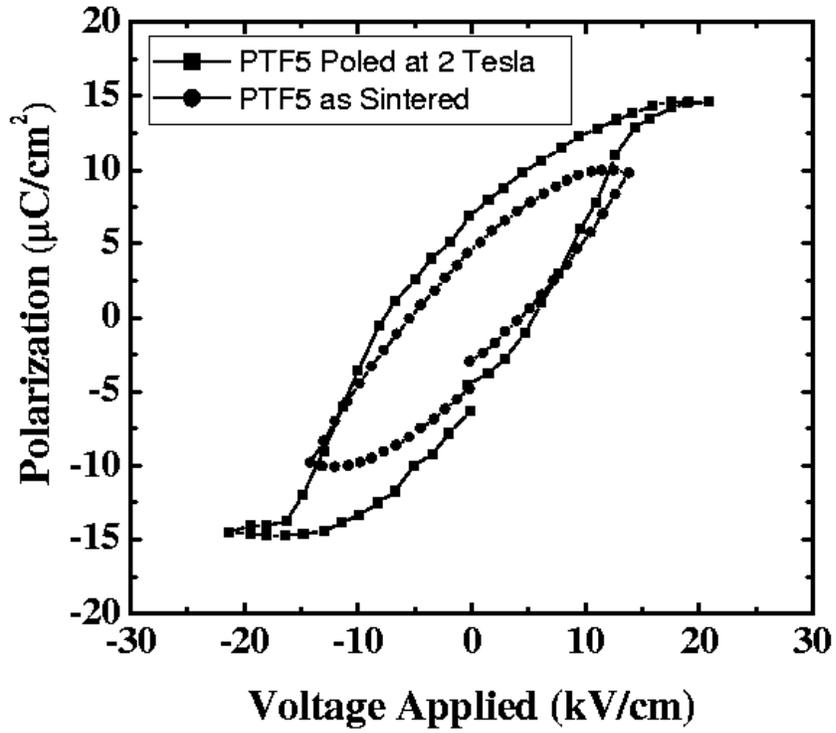

Figure 6. Palkar *et.al*

**Figure 6.** Ferroelectric Hysteresis Loop of Pb(Fe$_{0.5}$Ti$_{0.5}$)O$_3$ pellet sample, a) as sintered, b) poled at 2 Tesla of DC Magnetic Field.